\documentclass{elsart}
\usepackage{amssymb}
\usepackage{epsfig}

\newcommand{\ksl}{k \hskip -0.5em /}

\newcommand{\bi}[1]{\bibitem{#1}}

\newcommand{\bgi}{\begin{itemize}}
\newcommand{\eni}{\end{itemize}}

\newcommand{\bb}{\begin{thebibliography}}
\newcommand{\eb}{\end{thebibliography}}

\newcommand{\nn}{\nonumber \\}
\newcommand{\bwt}{\begin{widetext}}
\newcommand{\ewt}{\end{widetext}}
\newcommand{\be}{\begin{equation}}
\newcommand{\ee}{\end{equation}}
\newcommand{\ba}{\begin{eqnarray}}
\newcommand{\ea}{\end{eqnarray}}
\newcommand{\baa}{\begin{eqnarray*}}
\newcommand{\eaa}{\end{eqnarray*}}

\newfont{\fib}{cmfi10 at 10pt}

\newcommand{\Tr}{{\rm Tr}}

\renewcommand{\thefootnote}{\arabic{footnote}}

\newcommand{\bm}[1]{\mbox{\boldmath $#1$}}
\begin{document}

\begin{frontmatter}



\title{Chiral-odd Fragmentation Functions in Single 
Pion Inclusive Electroproduction
}


\author[label1]{Leonard P.  Gamberg}
\ead{lpg10@psu.edu}
\author[label2]{Dae Sung Hwang}
\ead{dshwang@sejong.ac.kr}
\author[label3,label4]{Karo A. Oganessyan}
\ead{kogan@mail.desy.de}
\address[label1]{Division of Science,
Penn State-Berks Lehigh Valley College, 
Reading, PA 19610, USA}
\address[label2]{Department of Physics, Sejong University, Seoul 143-747, Korea}
\address[label3]{INFN-Laboratori Nazionali di Frascati I-00044 Frascati, 
via Enrico Fermi 40, Italy
}
\address[label4]{ DESY, Deutsches Elektronen Synchrotron 
Notkestrasse 85, 22603 Hamburg, Germany 
}

\begin{abstract}
\baselineskip=4pt
We consider a sub-leading twist chiral-odd pion fragmentation function  
and explore  its contribution in single pion 
semi-inclusive electroproduction. We 
evaluate the single beam-spin azimuthal asymmetry $A_{LU}$ 
and the double spin asymmetry $A_{LT}$ in 
polarized electroproduction of pions from an unpolarized and 
transversely polarized nucleon respectively. 
The beam asymmetry is expressed as the product of chiral-odd, 
and $T$-odd and even distribution and fragmentation functions. 
The double spin asymmetry contains
information on the quark's transversity distribution.
In a quark diquark-spectator 
framework we estimate these asymmetries at $6\ {\rm GeV}$,
$ 12\, {\rm GeV}$, and $27.5\, {\rm GeV}$ energies.  

\end{abstract}

\end{frontmatter}

\vspace*{-1cm}
\section{Introduction}
\label{intro}
\vskip -0.75cm
One of the most interesting results in deep-inelastic spin physics has been
the discovery of a class of chirally odd quark distribution functions. That
which has garnered most attention is the leading twist
transversity distribution
$h_1$ which provides information on the quark transverse spin distribution
in a transversely polarized nucleon~\cite{RS,JJ91}. Chiral-odd distribution
functions are difficult to measure because they are suppressed in
inclusive deep inelastic scattering. However, when two hadrons participate
in the scattering process,
the nucleon's transversity can be accessed; for example,
in Drell-Yan scattering  with transversely polarized
protons~\cite{RS,RHIC}. 
Alternatively, 
transversity can be probed
in semi-inclusive deep inelastic scattering (SIDIS) where outgoing
hadrons are produced  in the current fragmentation region.
This process~\cite{cnpb93}
has been used as a filter to access transversity~\cite{HERMES}.
Here, the probability
for a transversely polarized quark to produce a pion is probed.
Other methods to probe transversity  involving
semi-inclusive production of $\Lambda$ hyperons, and of two pions have also
been  discussed in the literature~\cite{J96,JJT98,bacc00}.

Some time ago, Jaffe and Ji~\cite{JJ93}  suggested that
the nucleon's transversity could be probed
in polarized electroproduction of pions from a
transversely polarized nucleon. 
By comparison with the above mentioned Drell-Yan and  
single-spin asymmetry approaches, their proposal
is sub-leading in twist. However, since the measurement involves
merely one spinless particle in the final state it
proves to be an interesting approach to probe the effects
of higher twist in addition to providing a window into the
measurement of transversity.  The asymmetry characterizing
this process consists of a linear combination 
of two terms, one chiral-even and one chiral-odd.
To expose the the  chiral-odd effect of interest~\cite{J97}, 
the competing chiral-even mechanism must be subtracted away.

\begin{figure}[htb]
\centerline{\epsfxsize 6.0 cm {\epsfbox{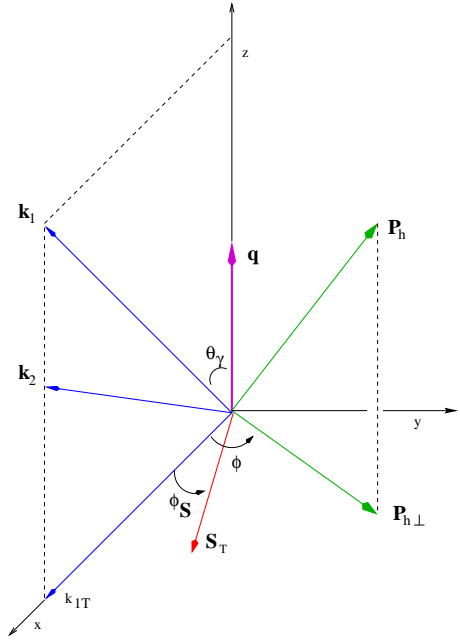}}}
\caption{\label{f1} \scriptsize The kinematics of SIDIS.}
\end{figure}

In this letter we will estimate the relative magnitudes
of these two contributions to the double spin asymmetry in
the quark-diquark spectator framework.
In doing so,  we explore the
sub-leading twist chirally odd pion fragmentation function  $E$ (in the
Ref.~\cite{JJ93} it was denoted by $\hat{e}_1$).
On the other hand, as it will be shown below, that chiral-odd fragmentation
function can show up also in the SIDIS beam spin asymmetry (BSA) in
addition to effects considered by Levelt and Mulders~\cite{LM}.
The interest in the BSA for pion electroproduction
in semi-inclusive deep inelastic scattering of longitudinally polarized
electrons off unpolarized nucleon resides in the fact that the beam
probes the antisymmetric part of the hadron tensor, which is particularly
sensitive to final state interactions.
In longitudinally polarized electron electromagnetic scattering, the BSA
shows up as a $\langle \sin\phi \rangle$
asymmetry for the produced hadron and is expressed as
\vskip -0.5cm
\ba
<\sin\phi>\,=\, \pm\, \langle 
\frac{{\mathsf s } \times \bm k_2 \cdot 
\bm P_{h\perp}}
{\vert {\mathsf s} \times \bm k_2 \vert 
\vert \bm P_{h\perp} \vert } \rangle,
\label{A1}
\ea
where $\mathsf s$ denotes the spin vector 
of the electron (the upper (lower)
sign for right (left) handed electrons), $\bm k_1$ ($\bm k_2$)
is three-dimensional vector of incoming (outgoing) electron momentum and
$\bm P_{h\perp}$ is the produced hadron's 
transverse momentum about virtual photon direction; $\phi$ is the azimuthal
angle of produced pions relative to the lepton scattering plane, and
$\phi_S$ is the azimuthal angle of the target polarization 
vector (Fig.~\ref{f1}).
This asymmetry is related to the
left-right asymmetry in the hadron momentum
distribution with respect to the electron scattering plane,
\vskip -0.5cm
\ba
A = \frac{\int_0^\pi d\phi d\sigma
- \int_\pi^{2\pi} d\phi d\sigma}{\int_0^\pi d\phi d\sigma
+ \int_\pi^{2\pi} d\phi d\sigma},
\label{A2}
\ea
which is $4/\pi$ times $<\sin\phi>$. Here $d\sigma$ is a 
shorthand notation for
$\scriptstyle d \sigma^{\vec{e}N \to e h X}/dx\,dy\,dz\,d^2P_{h\,\perp}$, and
$x$, $y$, and $z$ are the standard leptoproduction scaling variables~\cite{LM}.
As will be shown in this letter, the asymmetry is a superposition of 
$e\star H^\perp_1$, obtained in Ref.~\cite{LM}, and 
$h^\perp_1\star E$. A similar result has been presented very recently by Yuan ~\cite{Y03},
where the BSA has been phenomenologically studied in the 
extrema case of  $h^\perp_1\star E$. 
By contrast the  BSA has  been studied at the other extreme,  
$e\star H^\perp_1$~\cite{OADM,EGS} while in addition the bounds on this
asymmetry were considered in~\cite{lu_yang}. 
Further, the BSA has been considered 
solely from perturbative QCD effects 
at  second order in $\alpha_S$~\cite{HAG,OG,AG}. 
In the Ref.~\cite{AC} a dynamical 
model for BSA similar to that used for the single target spin 
asymmetries~\cite{bhs} has been proposed.     

This letter is organized as follows:
First we calculate the sub-leading twist chiral-odd
fragmentation function $E$, 
defined by Jaffe and Ji~\cite{JJ93}, following the approach
in~\cite{gamb_gold_ogan1,gamb_gold_ogan2}.  
Then we study its physical implications 
in filtering the nucleon's transversity properties
by considering the double-spin asymmetry, $A_{LT}$,
 in polarized electroproduction
of pions from a transversely polarized nucleon. 
This process contains  information on quark's transversity distribution.
We also consider the single beam-spin azimuthal asymmetry, $A_{LU}$.
We perform order of magnitude estimates 
of these asymmetries for HERMES 
and ongoing and upgraded JLAB energies~\cite{HERMES,white}.  With regard to
$A_{LT}$, we estimate that  the chiral-odd effect is small and apparently
its isolation from the chiral-even contribution would appear to be 
challenging measurement.  
\vspace*{-1cm}
\section{The $A_{LT}$ and $A_{LU}$ Asymmetries in the Spectator Framework}
\vskip -0.75cm
Here we focus on the the chirally odd transverse momentum dependent
distribution and fragmentation
functions, $e(x,\bm p_T)$ and $E(z, -z\bm k_T)$.
As mentioned in the introduction, the fragmentation function
$E(z, -z\bm k_T)$  arises when a longitudinal
polarized electron beam probes a transversely polarized nucleon. Quantitatively
this is represented in the joint product $h_1(x)\star  E(z)$ that arises
in the asymmetry $A_{LT}$ which is accompanied with the more commonly
investigated chiral-even combination, $g_{T}(x)\star  D_1(z)$.
On the other hand, the chirally odd distribution 
function $e(x)$ contributes in the 
combination $e(x)\star H_{1}^{\perp}(z)$ in the beam
asymmetry, where $H_{1}^{\perp}(z)$ correlates the
probability for a transversely polarized quark to fragment to a pion.
This term is accompanied with the complement $T$-odd
combination $h_1^{\perp}(x)\star E(z)$. The latter combination
provides an additional term that fuels the beam asymmetry.  The 
function $E(z,-z\bm k_T)$ is projected from
the fragmentation matrix $\Delta(k,P_h)$ using the identity operator
\vskip -0.5cm
\ba
\Delta^{\scriptscriptstyle \left[\mathbf 1\right]}(z, \bm k_T)
&=&\frac{1}{4z}\int dk^+ \Tr\left({\mathbf 1}\Delta(k,P_h)\right)
\Big|_{k^-=\frac{P^-_\pi}{z},\, \bm k_T}
=\frac{M_h}{P^-_\pi}E(z, z^2{\bm k_T^2}),
\label{delta}
\ea
where $\Delta(z,\ \bm k_T)$ is parameterized 
in terms of the relevant  fragmentation functions
\vskip -0.5cm
\ba
\Delta(z,\bm k_T) &=& \frac{1}{4}\Bigl\{ D_1(z,-z\bm k_T){\not{n_-}} +
H_1^{\perp}(z,-z\bm k_T)
\frac{{\sigma}^{\alpha\beta}k_{T\alpha}n_{-\beta}}{M_h}
+ \frac{M_h}{P_h^-}E(z,-z\bm k_T)
+\cdots \Bigr\}.
\nn
\label{d8}
\ea
Similarly, $e(x,\bm p_T)$ is projected from the
distribution matrix $\Phi(p,P)$,
\vskip -0.5cm
\ba
\Phi^{\scriptscriptstyle [{\mathbf 1} ]}(x,\bm p_T)
&=&\frac{1}{2}\int dk^- \Tr\left({\mathbf 1} \Phi(p,P)\right)
\Big|_{p^+=xP^+,\, \bm p_T}
=\frac{M}{P^+}e(x,{\bm p_T^2}) 
\label{phi}
\ea
which is parameterized as
\vskip -0.5cm
\ba
\Phi(x,\bm p_T) &=& {\frac{1}{2}}
\Bigl\{
f_1(x,\bm p_T){\not{n_+}} +
h_1^{\perp}(x,\bm p_T){{\sigma}^{\alpha\beta}p_{T\alpha}n_{+\beta}}{M} +
h_{1T}(x,\bm p_T)i\gamma_5 {\sigma}^{\alpha\beta}n_{+\alpha}S_{T\beta}
\nn &+&
h^\perp_{1s}(x,\bm p_T)\frac{i\gamma_5 {\sigma}^{\alpha\beta}
n_{+\alpha}p_{T\beta}}{M} + \frac{M}{P^+}\Bigl[e(x,\bm p_T)  +
g_T^\prime(x,\bm p_T)\gamma_5{\not{S_T}} 
\nn && \hspace{5.5cm} 
+ g_s^\perp(x,\bm p_T)\frac{\gamma_5{\not{p_T}}}{M}\Bigr]
\cdots \Bigr\}\ ,
\label{d7}
\ea
where we have used the shorthand naming convention~\cite{TM,BM}
\vskip -0.5cm
\ba
h^{\perp}_{1s} (x,\bm p_T) \equiv \lambda h^{\perp}_{1L}(x,\bm p_T)
+\frac{(\bm p_T\cdot S_T)}{M}h^{\perp}_{1T}(x,\bm p_T).
\ea
We  calculate these functions
in the spectator model framework~\cite{hood,JMR}.
To  address the $\log$ 
divergence~\cite{hood,bhs,ji,gold_gamb,bbh,gamb_gold_ogan}
that arises when calculating the moments of distribution and
fragmentation functions that 
appear in asymmetries,
we  introduce a Gaussian distribution
in the transverse momentum dependence of the
quark-spectator-pion  and quark-nucleon-spectator
vertices~\cite{hood,gamb_gold_ogan1,gamb_gold_ogan2}.  
This serves to smoothly
cutoff the integration in ${\bm k_T}$ which kinematically parameterizes
our knowledge of confining effects.
For the fragmentation vertex
we couple the on-shell spectator, as a quark
interacting with the produced pion (hereafter, $P_h=P_\pi$)
through the vertex function
\vskip -0.5cm
\ba
\langle 0\big| \psi(0)\big| P;X\rangle=
\left(\frac{i}{\ksl
 - m}\right)\Upsilon({\bm k_T^2})\, U({\scriptstyle k-P_\pi},s),\, \ 
{\rm where}\, \quad
\Upsilon({\bm k_T^2})=i \gamma_5 f_{qq\pi}
e^{-b^\prime {\bm k_T^2}}\ .
\label{fvert}
\ea
Here, $f_{qq\pi}(\equiv f)$  is the quark-pion
coupling and $k$ is the momentum of the off-shell quark,
$\bm k_T$ and $b^\prime=1/<{\bm k_T^2}>$, are the  intrinsic
transverse momentum and its inverse mean square respectively, and
$U(p,s)$ is the off-shell quark spinor.
A similar analysis applies to the quark-nucleon-spectator vertex as it relates
to the distribution function.
Using Eqs.~(\ref{delta},\ref{fvert}),
 the ${\bm k_T}$ integrated
chiral-odd twist-three fragmentation function is $E(z)$
\vskip -0.5cm
\ba
E(z)&=&\frac{m}{P^-}\frac{f^2}{4(2\pi)^2}\frac{1}{z}
\frac{\left(1-z\right)^2}{z^2}
\Bigg\{\frac{m_\pi^2}{\Lambda^\prime(0)}
-2b^\prime m_\pi^2
e^{2b^\prime\Lambda^\prime(0)}\Gamma(0,2b^\prime\Lambda^\prime(0))\Bigg\},
\ea
where $\Lambda^\prime(0)=\frac{1-z}{z^2}M_\pi^2
+ \frac{\mu^2}{z} -\frac{1-z}{z} m^2$.   The $T$-even distribution functions
$f_1(x)$, $h_1(x)$, and 
fragmentation functions,
$D_1(z)$,  $H_1^\perp(z)$ are detailed
in~\cite{gamb_gold_ogan1} and~\cite{gamb_gold_ogan2} (see also~\cite{BMY}).
Similarly using Eq.~(\ref{phi}),
 the  ${\bm p_T}$ integrated
chiral-odd distribution $e(x)$ function is
\vskip -0.5cm
\ba
e(x)&=&\frac{M}{4P^+}\frac{g^2}{(2\pi)^2}
\Bigg\{\frac{\left(1-x\right)\left(m+xM\right)\left(m+M\right)
-m^2\left(x+\frac{m}{M}\right)
+\Lambda(0)\left(1+\frac{m}{M}\right)}{\Lambda(0)}
\nn  &&\hspace{-0.35cm}
-\Bigg[2b\Bigg(\left(1-x\right)\left(m+xM\right)\left(m+M\right)
-m^2\left(x+\frac{m}{M}\right)
+\Lambda(0)\left(1+\frac{m}{M}\right)\Bigg)
\nn && \hspace{6.0cm}
+\left(1+\frac{m}{M}\right)\Bigg]
\times\,  e^{2b\Lambda(0)}\Gamma(0,2b\Lambda(0))\Bigg\}\, ,
\ea
where $g$ is the scalar diquark 
coupling~\cite{JMR}, 
$\Lambda(0)=(1-x)m^2 +x\lambda^2  -x(1-x)M^2$, while 
$M$ and $m$ are the nucleon and quark masses 
respectively. Choosing $<p_T^2> = {(0.4)}^2$ GeV$^2$ = $1/b$,  
yields good agreement~\cite{gamb_gold_ogan1,gamb_gold_ogan2} 
between $f_1(x)$ and the corresponding 
valence distribution of Ref.~\cite{GRV}.
Additionally the chiral-even polarized function 
is projected from Eq.~(\ref{d7}), 
\vskip -0.5cm
\ba
g_T(x)&=&\frac{M}{4P^+}\frac{g^2}{(2\pi)^2}
\Bigg\{\frac{\left(1-x\right)\left(m+xM\right)\left(m+M\right)
-\left(m^2-\Lambda(0)\right)\left(x+\frac{m}{M}\right)
}{\Lambda(0)}
\nn &&\hspace{-0.35cm}
-\left[2b\left(\left(1-x\right)\left(m+xM\right)\left(m+M\right)
-\left(m^2-\Lambda(0)\right)
\left(x+\frac{m}{M}\right)\right)
+\left(x+\frac{m}{M}\right)\right]
\nn && \hspace{8cm}
\times\,  e^{2b\Lambda(0)}\Gamma(0,2b\Lambda(0))\Bigg\}\, .
\ea
The distribution and fragmentation functions enter
 cross section for one-particle inclusive deep inelastic
scattering which is given by 
\vskip -0.5cm
\ba
\frac{d\sigma^{\ell+N\rightarrow \ell^\prime+h+X}}
{dxdydz d^2P_{h\perp}}=\frac{\pi\alpha^2 y}{2Q^4 z} L_{\mu\nu} 2M
{\mathcal W}^{\mu\nu}=\frac{2\,\pi \alpha^2}{Q^2 y} \sum_a e_a^2\sigma^a,
\label{CS}
\ea
where the
factorized~\cite{TM} hadronic tensor is
\begin{eqnarray}
2M{\mathcal W}^{\mu\nu}(q,P,P_h)&=&
\int d^2{\bm p}_Td^2{\bm k}_T\ {\delta}^2({\bm p}_T+{\bm q}_T-{\bm k}_T)
\nonumber
\\
&\times&
\frac{1}{4}\ {\rm Tr}\Big[\Phi(x_B,{\bm p}_T){\gamma}^{\mu}
\Delta(z_h,{\bm k}_T){\gamma}^{\nu}\Big]\, +\, 
\Bigl( q\leftrightarrow -q\ ,\ \mu \leftrightarrow \nu \Bigr)\  ,
\label{d15}
\end{eqnarray}
and $L_{\mu\nu}$ is the well-known lepton tensor.
To investigate 
the $\sin\phi$ BSA and $\phi$-independent $\sigma_{LT}$ SIDIS cross
section we keep only those terms producing 
 contributions to Eq.~(\ref{d15})  
\footnote{To avoid ambiguities, we will use the same notations
as in Ref. \cite{TM}. Also the terms proportional to the current quark
mass are neglected. }
\vskip -0.75cm
\ba
2M\,{\mathcal W}^{\mu\nu} &=&
2 z \int d^2{\bm p_T} d^2{\bm k_T}
\delta^2(\bm p_T - \frac{\bm P_{h\perp}}{\bm z}- \bm k_T)
\times\Biggl\{ -g_\perp^{\mu \nu} f_1 D_1
+ i\, \frac{2\,\hat{t}^{\{\mu} k_T^{\nu \}}}{Q} \frac{M}{M_h}\,
x\, e\, H^\perp_1
\nonumber\\ && \hspace{-1.0cm}
-\,  i\, \frac{2\,\hat{t}^{\{\mu} p_T^{\nu \}}}{Q} \frac{M_h}{M}\,
h^\perp_1\, \frac{E}{z}
+\,  i\,\frac{2\,\hat{t}^{[\mu} \epsilon_\perp^{\nu]\rho}p_{T \rho}}{Q}
\left [ \frac{({\bm p}_T\, \cdot {\bm S}_T)}{M}\,
\left ( x\, g^\perp_T\,D_1\,+\frac{M_h}{M} h^\perp_{1T}\,
\frac{E}{z}\, \right ) \right ]
\nonumber\\ && \hspace{-1.0cm}
 +i\frac{2M\,\hat{t}^{\, [\mu} \epsilon_\perp^{\nu ]\,\rho}\,
S_{T \rho}}{Q} \left [x\, \left ( g_T - \frac{{\bm p^2_T}}{2M^2}\,g^\perp_T
\right )\,D_1\,+\frac{M_h}{M}\,
\frac{E}{z}\,  \left (h_1\,-
\frac{{\bm p^2_T}}{2M^2}\,h^\perp_{1T} \right) \right ]
\Biggr\}.
\label{WMN}
\ea
Contracting the hadronic tensor with the helicity dependent part of the
leptonic tensor leads to the
 reduced cross sections which
 contribute to Eq.~(\ref{CS})
\vskip -0.5cm
\ba
\sigma^a &=& \int d^2{\bm p_T}\,d^2{\bm k_T}\,
z^2 \delta^2(\bm P_{h\perp} - z (\bm p_T - \bm k_T))
\times \Biggl\{
\left[1+(1-y)^2\right]\,f^a_1(x,\bm p_T^2)D^a_1(z,\,z^2 \bm k_T^2)
\nonumber \\ &-& 
 4 \lambda_e\,y\, \sqrt{1-y} \,\frac{1}{Q}\, \frac{M}{M_h} k_{Ty}\, x\,
e^a(x,\,\bm p_T^2) H^{\perp\,a}_1(z,\, z^2 \bm k_T^2)
\nonumber \\ &+& 
 4\lambda_e\,y\, \sqrt{1-y} \, \frac{1}{Q}\, \frac{M_h}{M} p_{Ty}\,
h^{\perp\,a}_1(x,\,\bm p_T^2) \frac{E^a(z,\, z^2 \bm k_T^2)}{z}
\nonumber \\ &+& 
  4 \lambda_e\,y\, \sqrt{1-y} \, \frac{1}{Q}\, S_{Tx}\,
\left ( M\,x\,g^a_T(x,\,\bm p_T^2) D^a_1(z,\, z^2 \bm k_T^2)
+ M_h\, h^a_1(x,\,\bm p_T^2) \frac{E^a(z,\, z^2 \bm k_T^2)}{z} \right )
\Biggr\} .
\nonumber \\
\label{SCS}
\ea
Here,  $k_{Ty}$ ($p_{Ty}$) denote the $y$ component of the
final (initial) parton transverse momentum vector and $S_{Tx}$ denotes 
the $x$ component of the nucleon's  polarization vector. We project the 
weighted differential cross section integrated 
over the transverse momentum of the produced hadron~\cite{KM,BM}
\be
{\langle W \rangle}_{AB}
= \int d^2P_{h\perp}\, W \frac{d\sigma^{\ell+N\rightarrow \ell^\prime+h+X}}
{dxdydz d^2P_{h\perp}},
\label{weight}
\ee
from Eq.~(\ref{SCS})  where $W = W(P_{h\perp},\phi,\phi_S)$. 
The subscripts $AB$ represent
the polarization of lepton and target hadron respectively,
$U$ for unpolarized, $L$ for longitudinally polarized
and $T$ for transversely polarized particles. 
From Eq.~(\ref{CS}) the relevant reduced cross sections terms
are~\footnote{Hereafter we omit $a$ assuming that the cross
section is given predominantly by scattering on the $u$-quark.}
\ba
\sigma_{UU} \equiv {\langle 1 \rangle}_{UU}
&=& \frac{[1+(1-y)^2]}{y}\,f_1(x) D_1(z),
\label{W1}
\\
\sigma_{LT} \equiv {\langle 1 \rangle}_{LT}
&=& \lambda_e \vert {\bm S_T} \vert\,\sqrt{1-y}\,
\frac{4}{Q}\, \cos\phi_S\, \Biggl [ M\,x\,g_T(x) D_1(z)
+ M_h\, h_1(x) \frac{E(z)}{z} \Biggr ],
\label{WLT}
\\
{\langle \vert P_{h\perp} \vert \sin\phi \rangle}_{LU}& =&
\lambda_e\,\sqrt{1-y}\, \frac{4}{Q}\, M M_h\,
\Biggl [ x\,e(x) z H^{\perp\,(1)}_1(z)\,+ h^{\perp\,(1)}_1(x)\,E(z)
\Biggr ],
\label{WLU}
\ea
where  $h_1(x)=h_{1T}(x)+h_{1T}^{\perp (1)}(x)$,  and
the weighted cross section contain 
the ${\bm p}^2_T$- and $ {\bm k}^2_T$-moments
of the distribution and fragmentation functions, 
\vskip -0.5cm
\ba
h_1^{\perp (1)}(x) \equiv  \int d^2 \bm p_T
\frac{ \bm p_T^2}{2M^2}
\,h_1^\perp(x, \bm p^2_T),\ {\rm and}\
H_1^{\perp (1)}(z)  \equiv  z^2\int d^2 \bm k_T
\frac{\bm k_T^2}{2M_h^2}
H_1^\perp(z,z^2 \bm k^2_T).
\ea
In turn, the asymmetries for which we will give an estimate are
the weighted integrals of a SIDIS cross section, Eq.~(\ref{weight}):
\be
A_{LU} \equiv A^{\vert P_{h\perp} \vert\,\sin\phi}_{LU} \equiv
\frac{\int d^2P_{h\perp} \vert P_{h\perp} \vert\,\sin \phi \left
(\sigma^{\leftarrow}-\sigma^{\rightarrow} \right )}
{\frac{1}{2}\int d^2P_{h\perp} \left( \sigma^{\leftarrow}
+\sigma^{\rightarrow} \right )} = 2\,\frac{{\langle \vert P_{h\perp}
\vert \sin\phi \rangle}_{LU}}{\sigma_{UU}},
\label{BSA}
\ee
\be
A_{LT} \equiv
\frac{\int d^2P_{h\perp}
\left(\sigma^{\leftarrow}(\phi_S)+\sigma^{\rightarrow}(\pi+\phi_S)
-\sigma^{\leftarrow}(\phi_S)-\sigma^{\rightarrow}(\pi+\phi_S) \right)}
{\int d^2P_{h\perp} \left ( \sigma^{\leftarrow}(\phi_S)
+\sigma^{\rightarrow}(\pi+\phi_S)
+\sigma^{\leftarrow}(\phi_S)+\sigma^{\rightarrow}(\pi+\phi_S) \right ) }
=\frac{\sigma_{LT}}{\sigma_{UU}}.
\label{LTASY}
\ee
Here $\sigma^{\leftarrow}(\phi_S), (\sigma^{\rightarrow}(\pi+\phi_S))$
denote the cross section with anti-parallel (parallel) polarization of the
beam and for a transversely polarized target. 
In numerical calculations we assume $100\%$ beam and target
polarization and $\cos\phi_S=1$.
\begin{figure}[htb]
\epsfxsize 7.5 cm {\epsfbox{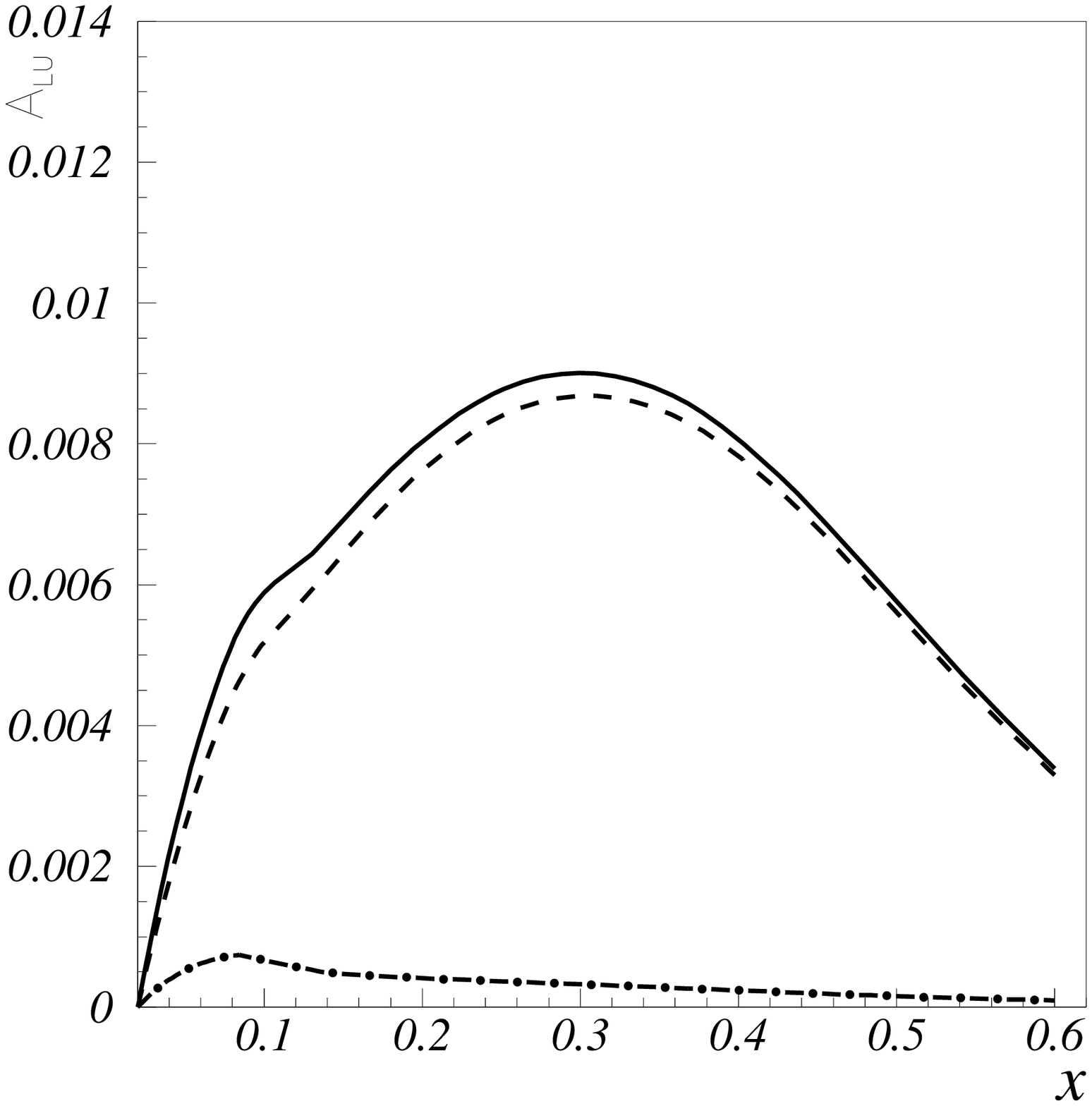}}
\epsfxsize 7.5 cm {\epsfbox{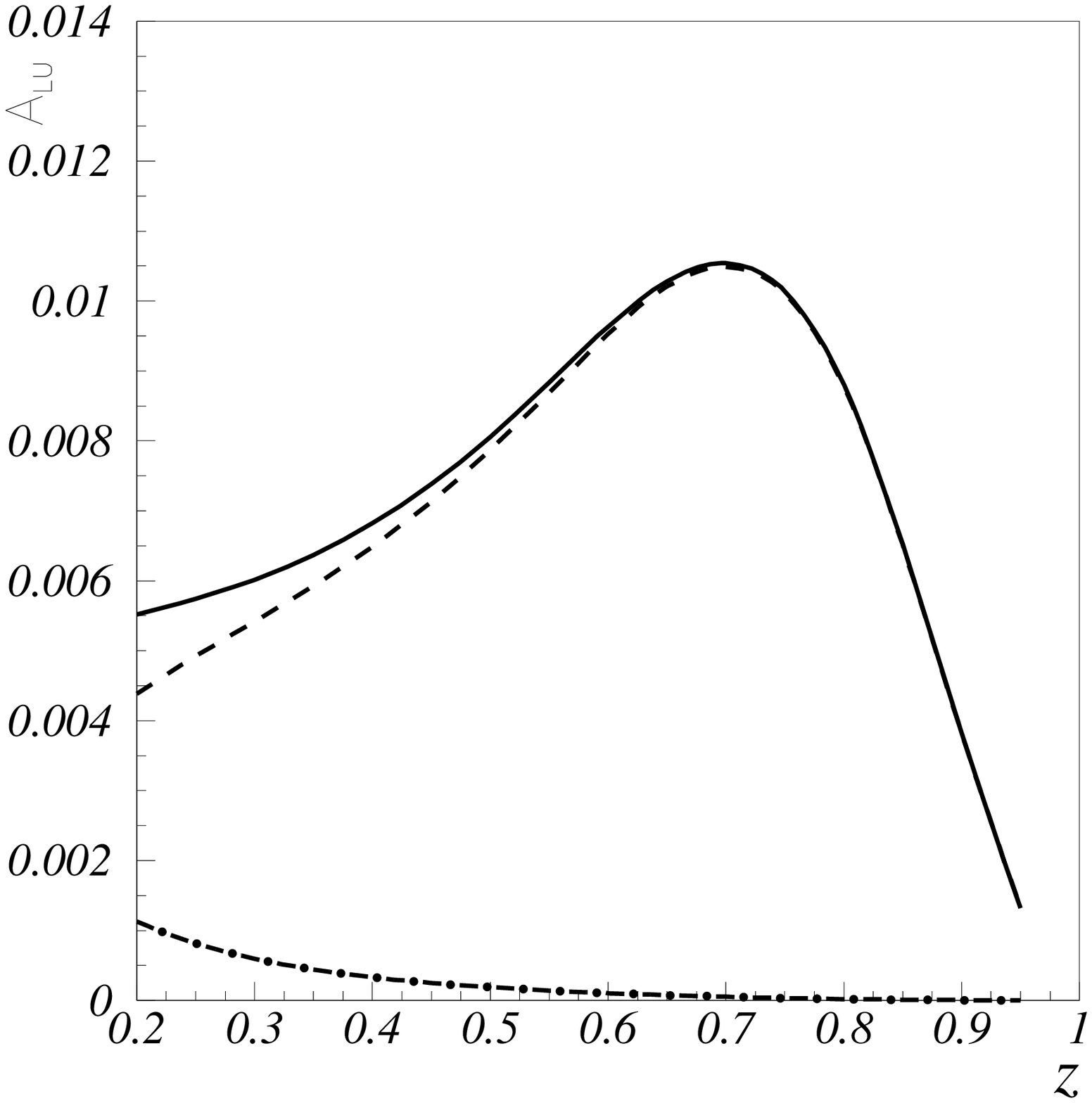}}
\caption{\label{f2}\scriptsize
$A_{LU}$ for $\pi^{+}$
production as a function of $x$ and $z$ at $27.5$~GeV energy. The dashed and
dot-dashed curves correspond to contribution of the first and second
terms of Eq.~(\ref{WLU}) respectively, and the full curve is the sum 
of the two.
}
\end{figure}

\begin{figure}[htb]
\epsfxsize 7.5 cm {\epsfbox{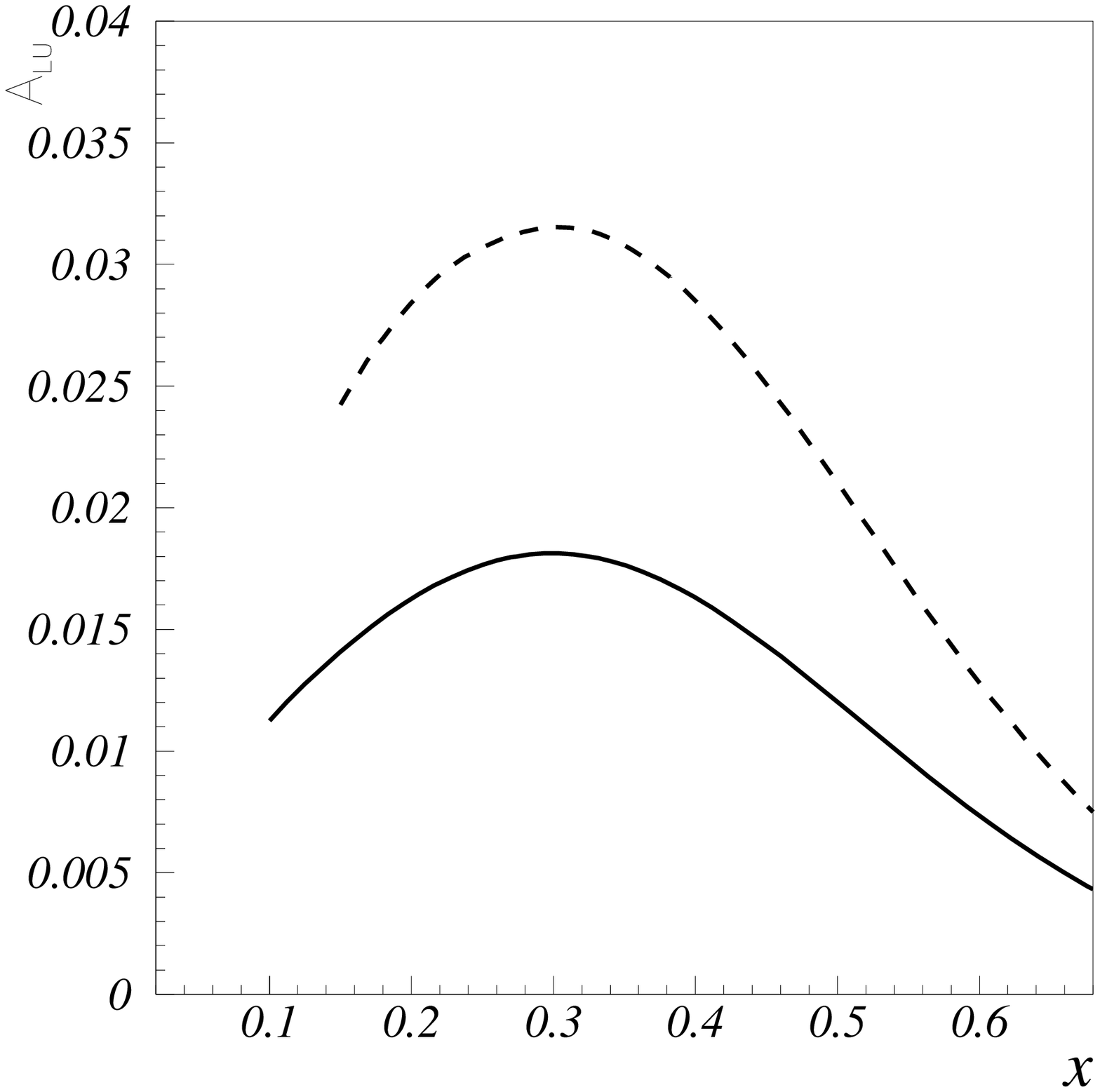}}
\epsfxsize 7.5 cm {\epsfbox{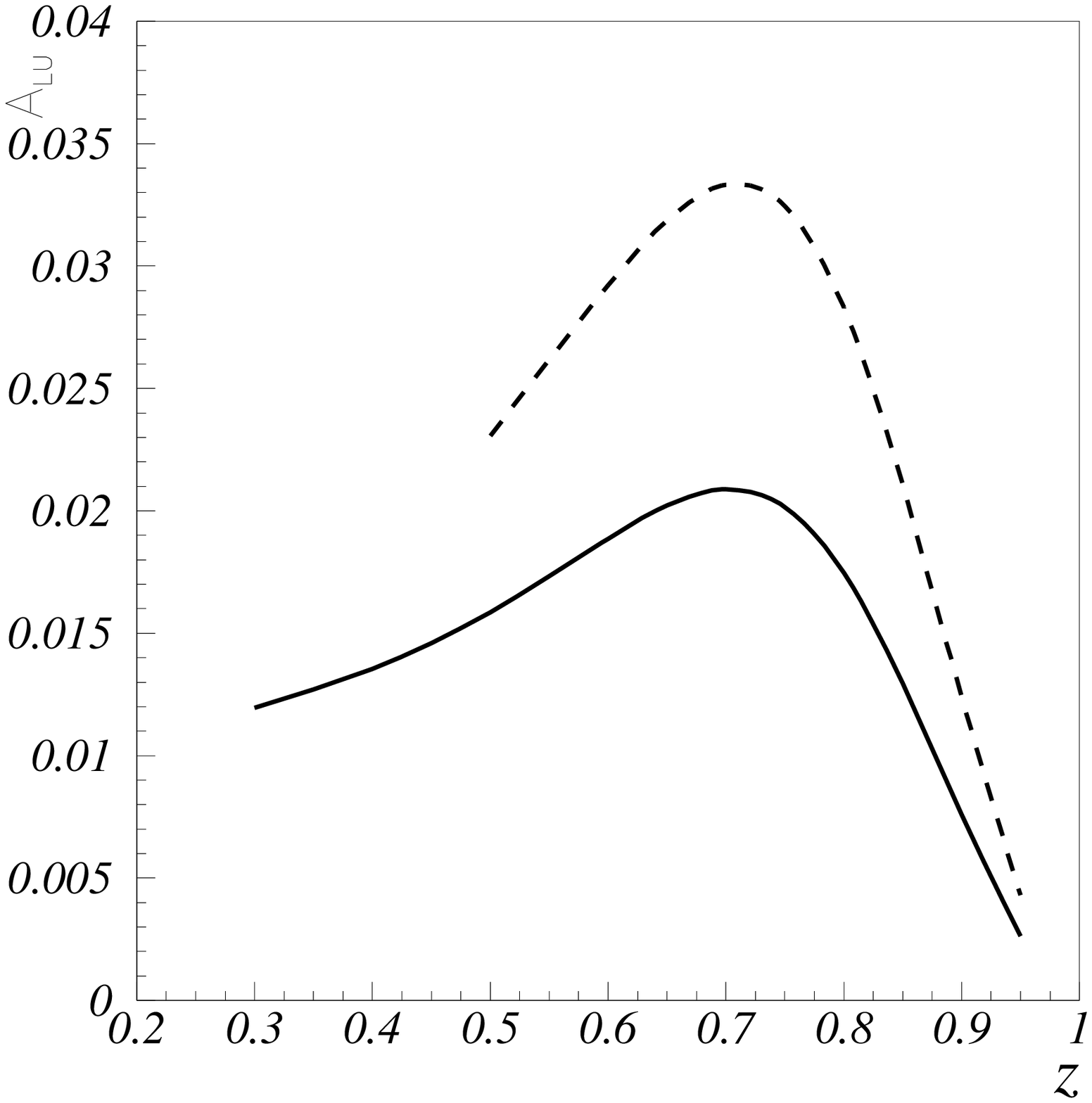}}
\caption{\label{f3}\scriptsize
$A_{LU}$ for $\pi^{+}$
production as a function of $x$ and $z$ at $6$~GeV and $12$~GeV energies.
The dashed curve corresponds to $6$~GeV and the full curve to $12$~GeV 
energies.
}
\end{figure}

\begin{figure}[htb]
\epsfxsize 7.5 cm {\epsfbox{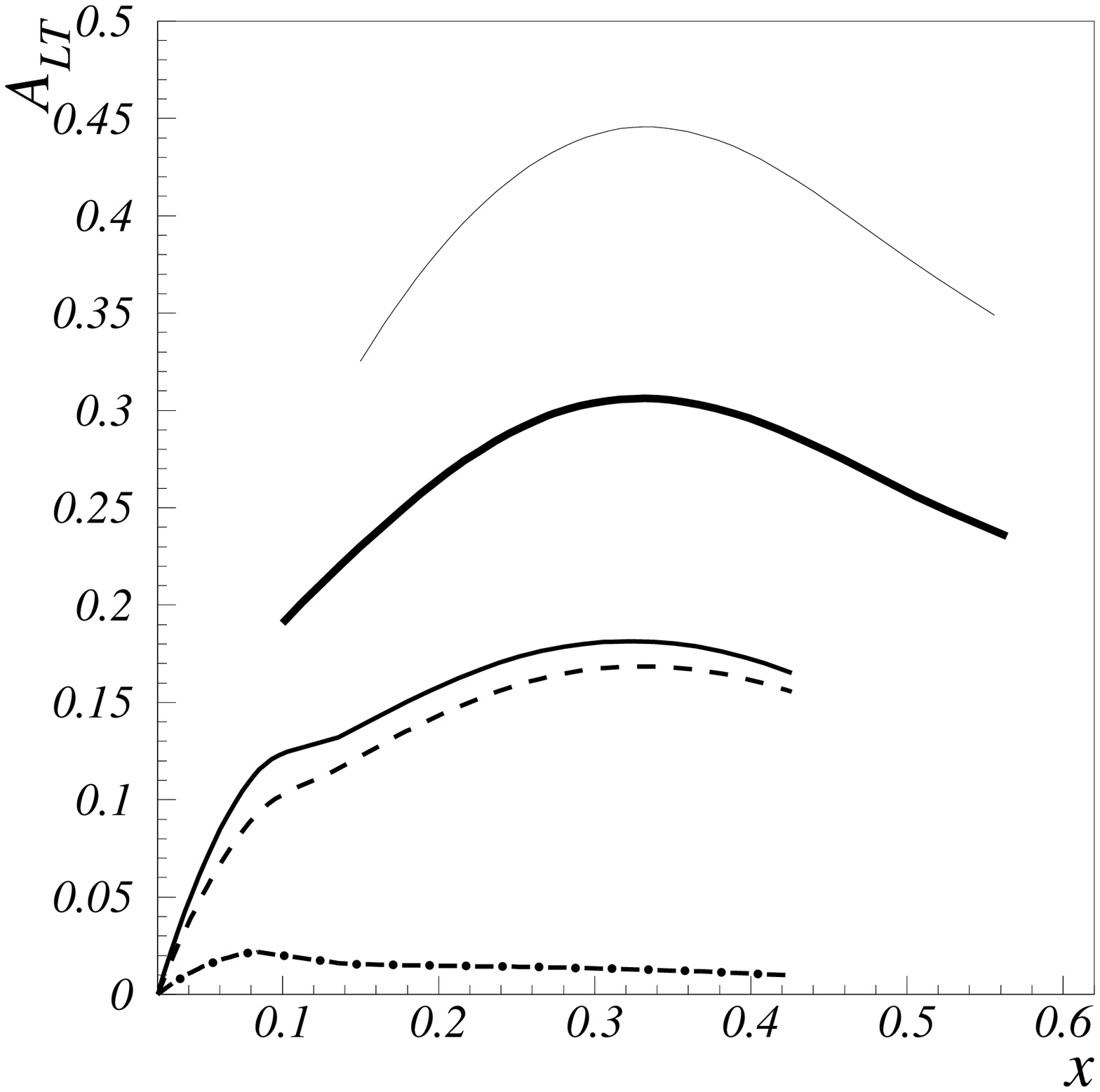}}
\epsfxsize 7.5 cm {\epsfbox{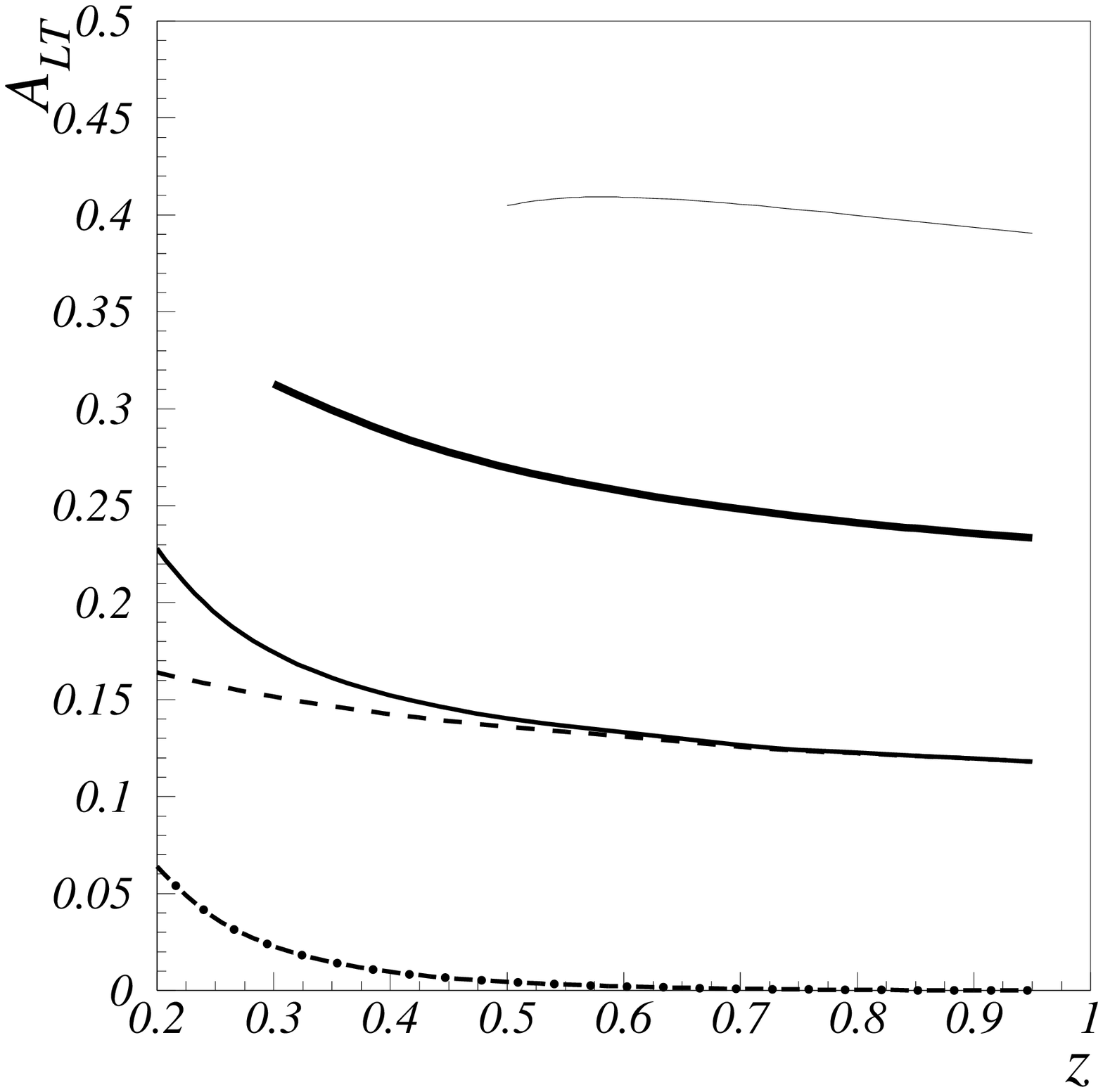}}
\caption{\label{f4}\scriptsize
$A_{LT}$ for $\pi^{+}$
production as a function of $x$ and $z$ at $27.5$~GeV energy.
 The dashed and dot-dashed curves correspond to the contributions of the two
terms of Eq.~(\ref{WLT}) respectively, and the full curve is the sum of those 
two. The thin curve corresponds to $6$~GeV and the thick curve to $12$~GeV
energies respectively.
}
\end{figure}
The curves in Figs.~\ref{f2},~\ref{f3}, and~\ref{f4}
are calculated at $27.5$~GeV, $12$~GeV,
and $6$~GeV beam energies by integrating over the kinematic ranges
corresponding to $0.1 \leq y \leq 0.85$, $Q^2 \geq 1$ GeV$^2$, and
$E_{\pi} \geq 2.0$~GeV. In Fig.~\ref{f2}, 
 the asymmetry $A_{LU}$ of Eq.~(\ref{BSA})
for $\pi^{+}$ production on a proton target is presented as a function of
$x$ and $z$.   The dashed and dot-dashed curves 
correspond to contribution of the 
two terms of Eq.~(\ref{WLU}) respectively, and the
full curve is the sum of the two. From Fig.~\ref{f2} one can see that
the contribution of the second term of Eq.~(\ref{BSA}),
$h^{\perp\,(1)}_1(x)\,E(z)$,  to  the beam spin
asymmetry is negligible whereas the first term, 
$e(x) H^{\perp\,(1)}_1(z)$, dominates.
This is to be contrasted with the result obtained 
in the Ref.~\cite{Y03}, where the $z$ dependence of the $A^{\sin\phi}_{LU}$
results solely from the ratio of $E(z)$ to $D_1(z)$ calculated in the chiral
quark model~\cite{JZ94}. In Fig.~\ref{f3}, 
the BSA, $A_{LU}$, is presented. The dashed curve 
corresponds to the full asymmetry at $6$~GeV beam 
energy and similarly,  the full curve corresponds to 
$12$~GeV beam energy.
It is apparent that decreasing the beam energy results in an
increasing BSA, which is consistent with it being a twist-three effect, 
suppressed by  ${\mathcal O}{(1/Q)}$. 
In Fig.~\ref{f4}, the asymmetry $A_{LT}(x)$ of Eq.~(\ref{LTASY}) for $\pi^{+}$
production as a function of Bjorken $x$ and $z$  is presented.  
The dashed and dot-dashed curves correspond to the contribution of the 
chiral-even-even and chiral-odd-odd  terms of Eq.~(\ref{WLT}), 
respectively, and the full curve is the sum of
the two. The thin and thick curves correspond to $6$~GeV and $12$~GeV
beam energies, showing  the total asymmetry. The contribution of the
term responsible for transversity in $A_{LT}$ is suppressed due to the
pion mass and the factor $1/z$.  From Fig.~\ref{f4} one can conclude that the
isolation  of the chiral-odd effect
containing information on quarks transversity from the
 term, $g_T D_1$, would to present a challenging 
measurement.
\vspace*{-1cm}
\section{Conclusion}
\vskip -0.75cm
The double transverse spin asymmetry which proves to be an interesting
observable to probe the effects of higher twist in addition to providing
a window into the measurement of transversity
has been considered in the quark-scalar diquark
framework~\cite{gamb_gold_ogan1,gamb_gold_ogan2}. 
In this connection, we have explored the twist-three chiral-odd 
pion fragmentation function and subsequently 
estimated the double-spin
asymmetry with longitudinally polarized electrons scattered on transversely
polarized nucleons. This asymmetry  contains 
the product of a chirally odd twist-two transversity distribution 
and a twist-three fragmentation function. At HERMES~\cite{HERMES} and
 ongoing  and  upgraded JLAB~\cite{white}
energies this chiral-odd effect is estimated to be fairly small which makes 
its isolation from the chiral-even mechanism challenging.  In 
addition the beam spin azimuthal asymmetry, which also
contains this sub-leading twist
chirally odd fragmentation function, has been calculated
for  HERMES and JLAB kinematics.  It is shown that in the
simple quark-diquark model the effects of the
twist-three chirally odd
fragmentation are suppressed. Consequently, the measurements of BSA can
provide   valuable information on the leading $T$-odd fragmentation function,
$H^\perp_1$, a favored candidate for filtering the transversity properties
of the nucleon.

The approach presented in this letter takes into account only up quarks.
However the inclusion of axial-vector diquarks may essentially affect 
the asymmetries~\cite{BSY03}. The extension
of our results for down quarks and estimates of BSA and double spin
asymmetries for $\pi^-$ and $\pi^0$ is a subject of further studies.
\begin{flushleft}
{\bf Acknowledgments}\\
We thank Gary Goldstein for valuable discussions.
\end{flushleft}

\end{document}